\documentclass{aastex62}
\usepackage[flushleft]{threeparttable}
\usepackage{CJK}
\usepackage{overpic}

\graphicspath{{./}{figures/}}

\submitjournal{ApJ}

\shorttitle{Gamma-rays from Flare Stars}
\shortauthors{Song, Paglione}

\begin{document}
\begin{CJK*}{UTF8}{gbsn}

\title{A Stacking Search for Gamma-ray Emission from Nearby Flare Stars and the Periodic Source TVLM 513-46546}

\correspondingauthor{Yuzhe Song}
\email{ysong2@gc.cuny.edu}

\author[0000-0002-2080-9232]{Yuzhe Song(宋宇哲)}
\affiliation{the Physics Program, the Graduate Center, City University of New York, 365 Fifth Ave., New York, NY 10016, USA}
\affiliation{Department of Earth \& Physical Sciences, York
College, City University of New York, 94-20 Guy R. Brewer Blvd.,
Jamaica, NY 11451, USA}
\affiliation{Department of Astrophysics, American Museum of
Natural History, Central Park West at 79th Street, New York, NY 10024, USA}

\author[0000-0001-9139-0945]{Timothy A. D. Paglione}
\affiliation{Department of Physics, the Graduate Center, City University of New York, 365 Fifth Ave., New York, NY 10016, USA}
\affiliation{Department of Earth \& Physical Sciences, York
College, City University of New York, 94-20 Guy R. Brewer Blvd.,
Jamaica, NY 11451, USA}
\affiliation{Department of Astrophysics, American Museum of
Natural History, Central Park West at 79th Street, New York, NY 10024, USA}

\begin{abstract} 
So far, the Sun is the only isolated main sequence star detected in gamma-rays, particularly during powerful flares. Young ultracool dwarfs are far more active so they are also plausible gamma-ray sources. We performed a spatial stack of 97 of the nearest X-ray and radio flare stars to search for GeV emission using nearly 12 years of data from the {\it Fermi Gamma-ray Space Telescope}. The stacked residual maps showed no significant signal. Modeling the upper limits indicates a peak stellar flux at least a factor of 7 below the noise level. We also analyzed the phase-folded light curve of the rapidly rotating radio star TVLM 513-46546, report a tentative (TS = 30) pulsed signal, and refine its period. We examine the possibility of a false positive signal by analyzing nearby {\it Fermi} catalog sources and test fields, and by repeating the analysis using different periods. No other periodic signals are found, despite clear detections of the catalog sources, and the TS value for TVLM 513 increases systematically to the optimal period. The putative gamma-ray signal is nearly in phase with the optical peak, and out of phase with the radio pulses by $0.4 \pm 0.05$ rotations. These results argue for emission from relativistic protons streaming down flux tubes towards the photospheric active regions. The protons colliding with the atmosphere create neutral pions that decay into gamma-ray photons. This would be the first detection of a normal, isolated star in gamma-rays, and the strongest evidence yet for proton acceleration in stellar magnetospheres.

\end{abstract}


\object{TVLM 513-46546} 

\section{Introduction} 
\label{sec:intro}

Young, low mass stars can be exceptionally active, often exhibiting flares and outbursts exceeding the Sun's activity and flare energy output despite their otherwise cool temperatures and low quiescent luminosities. This magnetic activity is particularly enigmatic in very low mass stars, which should have fully convective interiors, and therefore a global dynamo distinct from the Sun's. Solar and stellar activity is evident as flux variations seen from radio to x-ray, in continuum and spectral lines, and even gamma-rays in the case of the Sun \citep{Omodei2018}. GeV emission is notable in that it is a fairly unambiguous sign of the acceleration of protons and ions -- not just electrons -- which generally requires strong shocks accompanying powerful events such as coronal mass ejections. Such extreme events could prove devastating to the potential habitability of planets orbiting these stars. Since M dwarfs are the most abundant stellar type in the Milky Way, violent stellar activity in young systems, if common, could severely limit the prospect for life throughout the Galaxy.

GeV gamma-ray emission generally requires the creation of a non-thermal population of relativistic protons which interacts with ambient gas to generate pions. A number of flare stars show clear evidence of acceleration of at least electrons by their radio emission. The radio emission is notably correlated with the thermal emission from heated atmospheric gas seen in x-rays \citep{BenzGuedel2010}. Not limited to flares, similarly linked thermal and nonthermal emission mechanisms are associated with the auroral emission from rapidly rotating dwarf stars as well as giant planets \citep{Pineda2017}. \citet{OhmHoischen2018} estimated the expected MeV-TeV emission from the most energetic superflares observed from nearby stars to date. Based on the 2014 superflares from the interacting binary DG CVn, they predicted fluxes just out of the reach of $Fermi$ and current Cherenkov telescopes. Attempts to detect the DG CVn superflare in the MeV-TeV range resulted in upper limits \citep{loh2017, mirzoyan2014} factors of a few to several above the predictions. Recent flare surveys using TESS, Kepler, and Evryscope observations have quantified the flare rates and flare frequency distributions indicating a significant population of potential superflare sources in the solar neighborhood \citep{Loyd2018, YangLiu2019, howard2019, Gunther2019}. 
The abundance of young, cool dwarfs within a few dozen pc displaying frequent activity motivates a stacking survey to aggregate any signal. Further, some radio stars show periodic emission and their signal may be temporally stacked to elicit a detection in the phase-folded light curve. Here we attempt to accumulate any $\gamma$-ray signal from active stars with a stacking survey of the nearest and most x-ray- and radio-bright dwarf stars using nearly 12 years of $Fermi$-LAT data. One of the sources, TVLM 513-46546 (hereafter TVLM 513), exhibits periodic emission in the radio and optical and has a rotation period known precisely enough to allow a temporal stack of the signal.


\section{Observations} \label{sec:observations}

The {\it Fermi Gamma-ray Space Telescope} is a very natural instrument for such a long-term stacking survey. In near low-Earth orbit since 2008 and in practically continual survey mode, the $Fermi$ Large Area Telescope (LAT) has compiled exceptional exposure of the gamma-ray sky above 100 MeV unlike any project before. The third revision of the PASS8 data, P8R3, released on Nov 26, 2018 was used in this study, along with the most recent source catalog \citep[the 4FGL,][]{fermi2019} and diffuse background models \citep{abdo2009}. The $Fermi$ Science tools conda distribution version 1.2.23 were used for data analysis in this work\footnote{https://fermi.gsfc.nasa.gov/ssc/data/analysis/}. For all the sources in the spatial stack and the the temporal analysis of TVLM 513, we used data between 100 MeV and 300 GeV, from MET = 239557417s to MET = 608082000 (week 9 to 618), and within a region of interest (ROI) of $10\degr$ radius.

For spatial stacking, the young M dwarf sample consists of known x-ray and radio emitting stars above Galactic latitudes of roughly $|b| > 20\degr$ and a few other very well-known, isolated flare stars tabulated in \citet{reid2005}. The Galactic plane is a significant gamma-ray source, so targeting high latitudes is essential for reducing any systematic effects from modeling the background. X-ray-emitting M dwarfs were sampled from the catalog of nearby, low mass stars by \citet{shkolnik2009}, which were spectroscopically determined to be young and therefore presumably magnetically active. The list was sorted by the X-ray flux measured with the ROSAT All-Sky Survey. Radio-emitting M dwarfs were taken from the survey by \citet{mclean2012}. The stars with undetected quiescent radio emission were removed unless they had detectable radio flares. Binary systems were avoided for this study to focus only on stellar activity. A total of 97 stars were used in the spatial stacking survey.

The source used for temporal analysis in this study is the periodic M9 dwarf TVLM 513 \citep{henry1993}. Periodicity was first detected in the radio and confirmed in the optical with a consistent period, indicating the rotational origin of the signal. Its optical period is $P_0 = 1.95958 \pm 0.00005$ h \citep{harding2013}. 


\section{Stacking Survey of the Flare Satrs} \label{sec:methods}

\subsection{Methods and Results} \label{subsec:sstack}

Spatial stacking starts with the standard likelihood analysis using the $Fermi$ Science Tools. The data were filtered using {\tt gtselect} to define the ROI, energy range, and set a zenith cut of $90\degr$ to avoid the limb of the Earth and to account for the PSF of the LAT at lower energies. {\tt gtmktime} was used with {\tt (DATA\_QUAL > 0)\&\&(LAT\_CONFIG==1)} criteria to filter data within the good time intervals. Version v01r04 of make4FGLxml.py was used to create the model file of each ROI, which automatically applied a $10\degr$ padding around the ROI with the spectral parameters fixed. The binned counts cube was created with {\tt gtbin}, and {\tt gtltcube} and {\tt gtexpcube2} were used to calculate the exposure map of the sky before the likelihood analysis was completed using {\tt gtlike}. The signifcance of any source is judged by its test statistic, TS $=2 \log{{\cal L}/{\cal L}_0}$, a comparison of the log likelihoods of a source model and the null hypothesis (no source, ${\cal L}_0$). TS maps of each ROI were calculated with {\tt gttsmap} after fixing all the free parameters in the model file. Within each map, at any position with TS values larger than 25, we added a point source with a power law spectrum to the model file, and ran {\tt gtlike} once more. A model map was created using {\tt gtmodel} and a counts map was created using {\tt gtbin} with the CMAP option and $0.1\degr$ pixels. Using {\tt farith} from HEASoft, a residual map of the region was calculated by subtracting the model map from the counts map. The signal and noise counts are derived from the residual map. The residual maps were weighted to a standard distance of 10 pc.
The signal count is calculated from the root-mean-square of all the residual values of the 317 pixels within $1\degr$ of the center of ROI. The noise count is calculated from the root-mean-square of all the residual values of the 317 pixels within an annulus between $1\degr$ and $1.4\degr$ around the center of the ROI. 

\begin{figure}
\epsscale{0.7}
\plotone{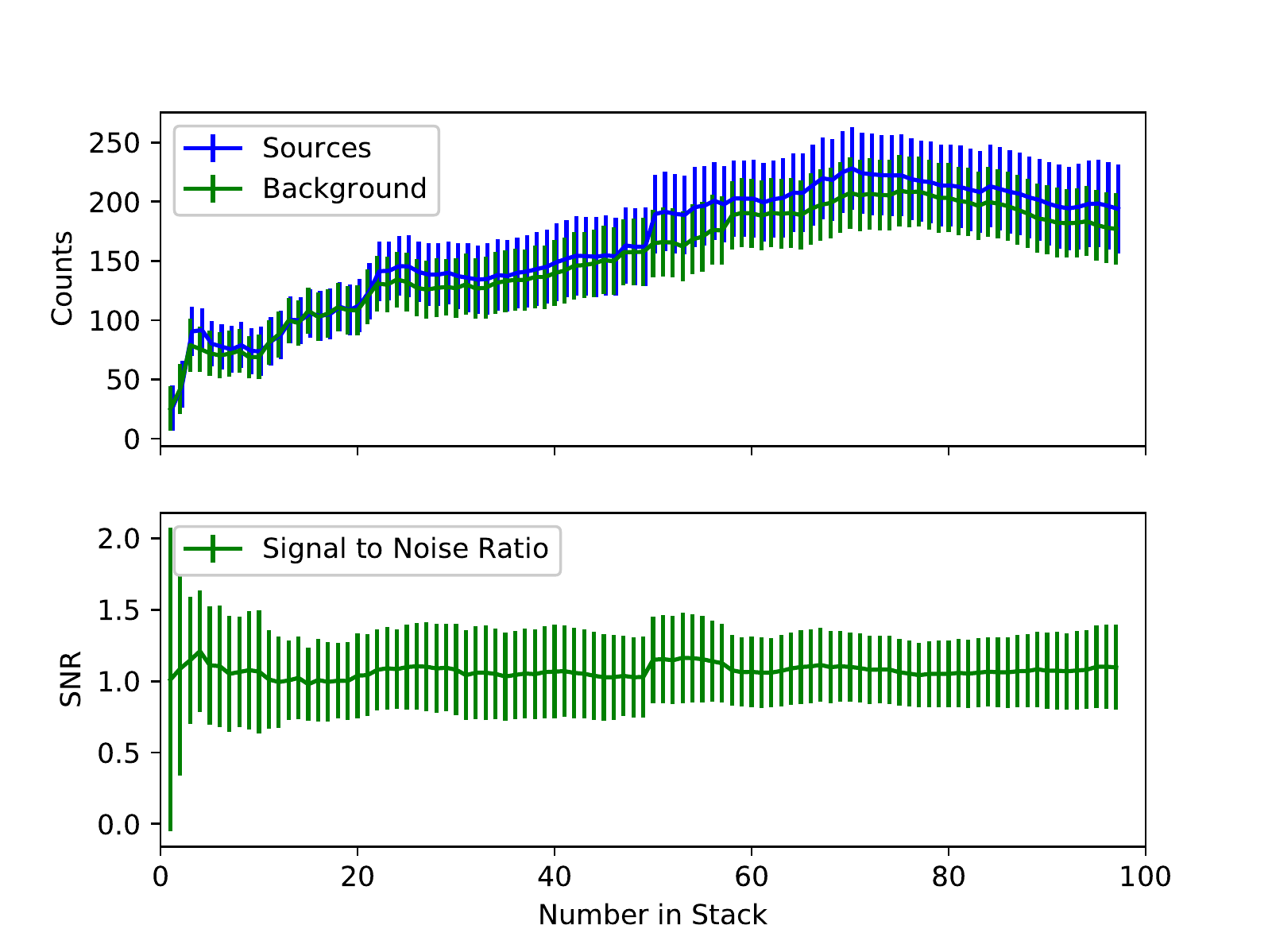}
     \caption{Results of spatial stack. The upper panel shows the stacked counts of the sources and the background versus the number of sources stacked, blue being the counts from sources and green being the noise counts. The curve of counts from sources was shifted 0.2 in number to the right to make error bars visible on both curves. The lower panel is the signal-to-noise ratio (SNR) versus the number of sources stacked. The error bars are estimated from a bootstrap resampling of the stack.}
    \label{fig:spatialresults}
\end{figure}

Following the methods above, 97 residual maps of the selected sources were created. Stacking was achieved by adding up any desired number of residual maps again with {\tt farith}, with signal and noise counts calculated from the resulting stacks. The stacking results are presented in Fig.~\ref{fig:spatialresults}. The sources were sorted by brightness with the X-ray survey first, followed by the radio stars, then the historical flaring stars. The uncertainties in Fig.~\ref{fig:spatialresults} were calculated from 100 bootstrap realizations of resampling the stack. Since the signal-to-noise level remains near unity, no $\gamma$-ray emission from these flaring stars was detected.

\subsection{Stack Model}
\label{subsec:smodel}

To test the expected stacked signal from the 97 flare stars and judge the limits imposed by the non-detection, we constructed a simple flare star model. We assume a Poisson noise level with an average of 18 counts, consistent with the individual residual counts maps, and create 97 counts maps with the same pixel size and map area as the survey. We model the stars in the stack as circular Gaussian counts distributions with full-width at half-maximum of $0.5 \degr$, sorted by decreasing flux. We assume the stellar fluxes follow a power law distribution with an index $\alpha$ estimated from the cumulative flare frequency distributions of Evryscope and other works \citep{howard2019}. This power law index is in the range of $-0.5$ to $-1.0$. We vary the peak signal-to-noise ratio using Poisson count rates to test the detection significance.

\begin{figure}
      \epsscale{0.7}
     \plotone{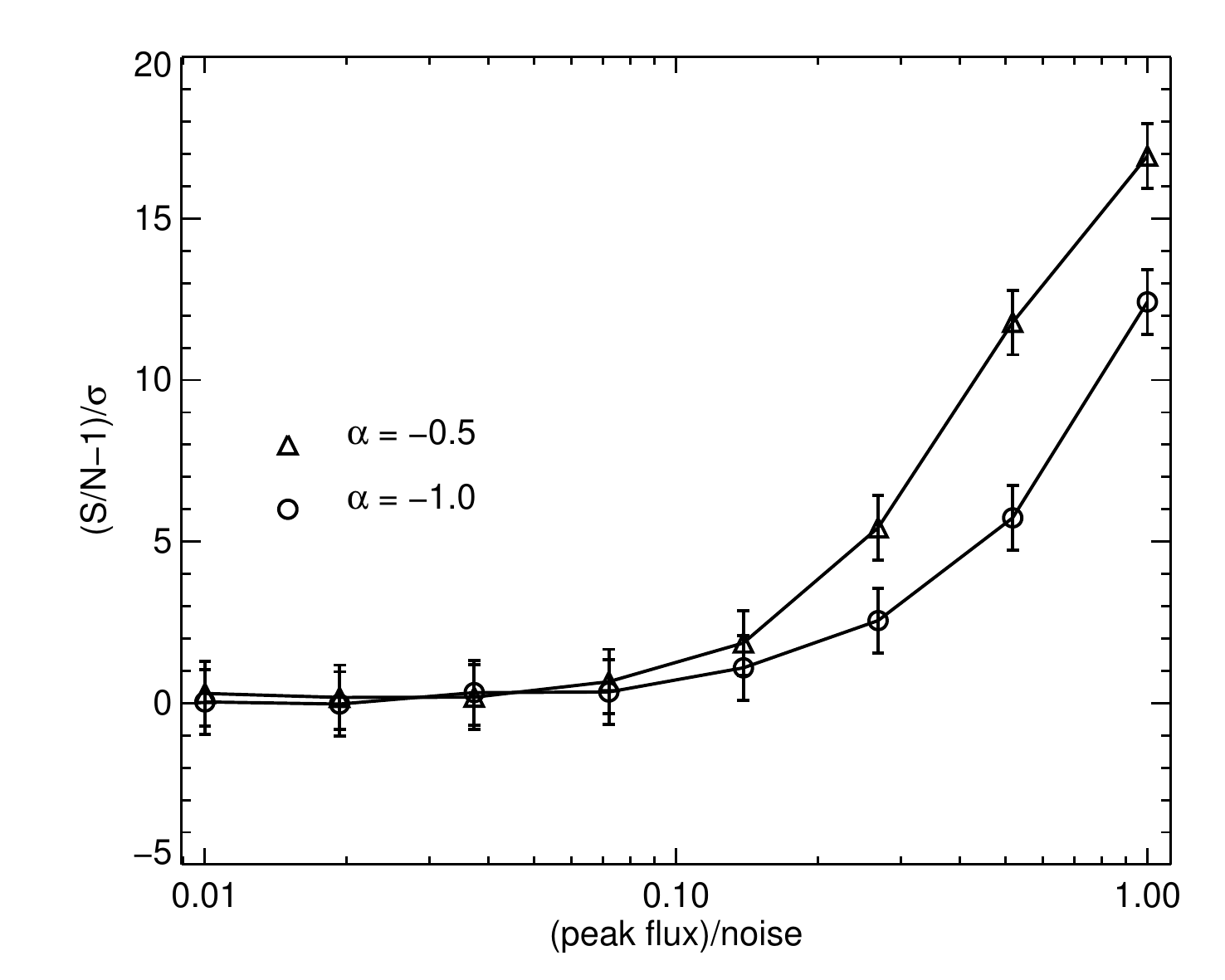}
     \caption{Modeled detection significance in the stack for two populations of flare stars with different power law indices, $\alpha$, for the cumulative flare frequency distribution. Points indicate the highest mean significance from 100 stacking trials, and the errorbars indicate the standard deviation in the stacked significance.}
     \label{fig:stackmodel}
\end{figure}

We estimate the expected detection significance by comparing the signal and noise counts as in our data analysis. Fig.~\ref{fig:stackmodel} shows the results of 100 trials for steep and shallow flux distributions. For $\alpha = -0.5$, a stacked detection becomes apparent even if the brightest star (or equivalently, the brightest flare) has a flux 7-10 times below the noise level. The detection threshold is slightly higher for steeper distributions, naturally, since nearly all of the stacked flux comes from the brightest star. The stack significance grows for a shallow flare frequency distribution, while for $\alpha=-1.0$, the significance is highest at the beginning of the stack.

\begin{figure}
    \epsscale{0.7}
    \plotone{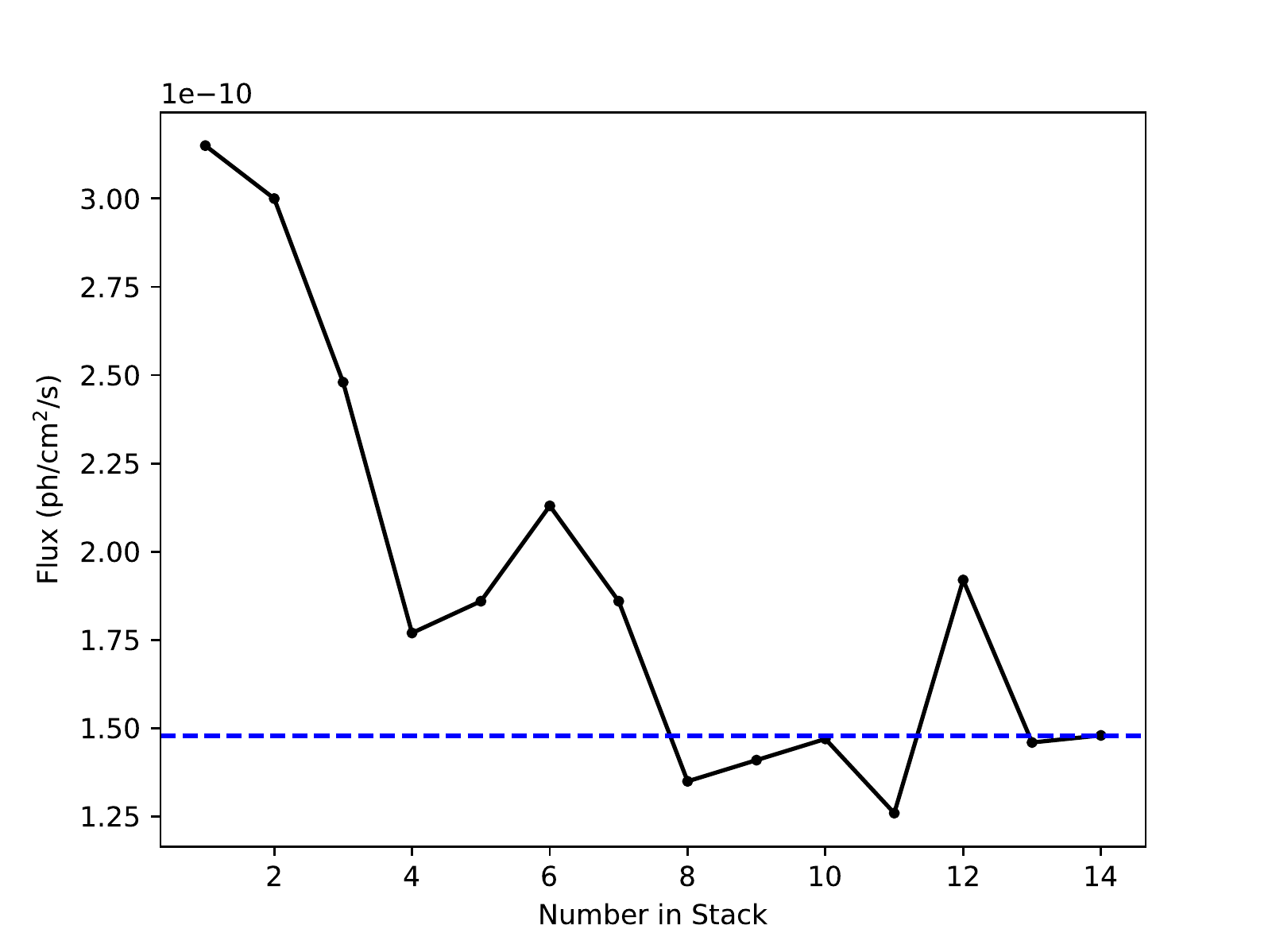}
    \caption{95\%\ confidence upper limit versus the number of ROIs stacked. The photon index chosen for the stars is 2.57, from the results of temporal analysis of TVLM 513 as in Sec.~\ref{subsec:tresults}. The blue dashed line is the averaged upper limit excluding the first six stars.}
    \label{fig:spatial_uls}
\end{figure}

We estimate the 95\%\ confidence flux upper limits following the procedure of \citet{Huber2012}, and focusing on just the brightest radio flare stars in the survey. Fig.~\ref{fig:spatial_uls} indicates that we reach a sensitivity floor of $1.5\times 10^{-10}\ {\rm ph\ cm}^{-2}\ {\rm s}^{-1}$ after stacking the first 14 stars. This limit is well below the LAT sensitivity\footnote{https://www.slac.stanford.edu/exp/glast/groups/canda/lat\_Performance.htm} of about $10^{-9}\ {\rm ph\ cm}^{-2}\ {\rm s}^{-1}$ for the flux above 100 MeV of a high latitude point source with a power law spectrum and photon index of 2. 

\section{Temporal Analysis of TVLM 513}\label{sec:temporal}
\subsection{Methods} 
\label{subsec:temporalanalysis}

Some radio stars emit periodically by bringing emission regions into the field of view as they rotate \citep[e.\ g., ][]{Yu2011, Lynch2015}
As anticipated for photometry of any individual source, the likelihood analysis for the periodic radio star TVLM 513 did not indicate a significant detection within the ROI. In fact, the upper limit of its flux is $9.2 \times 10^{-9}$ ph cm$^{-2}$ s$^{-1}$, consistent with the previous null detection in \S~\ref{subsec:sstack}. We investigated a number of methods to test for any emission at the known rotational period.

On the desired ROI within the observation time range, we attempted an aperture photometry analysis, using {\tt gtbin} with the LC option and a bin size of 705.4488 s. We also used the rotational information in \citet{wols2014} to construct the ephemeris of the star and
a phase-folded light curve using {\tt gtpphase}. The obtained light curves with both methods are dominated by the random variability of a relatively bright BL Lac near the center of the ROI, 4FGL J1501.0+2238 (hereafter 4FGL 1501) since it contributes most of the counts within the ROI. We tested the data with the commonly utilized Lomb-Scargle periodogram \citep{lomb1976,scargle1982}, $Z_m^2$ test \citep{beran1969,buccheri1983} and H-test \citep{dejager1989} as well. However, with the presence of 4FGL 1501 and the modulation of the spacecraft, any sub-threshold periodic or pulsed signal from TVLM 513 cannot be detected with these methods. The LS periodogram results only showed the 1.6 hr orbital period of the spacecraft, while the $Z_m^2$ test and H-test had low test statistics given this ephemeris. These tests proved effective on the Crab pulsar, however, given its bright, isolated nature. With its period of 33 ms, it should appear as a constant source in time bins of $\sim 12$ minutes. Using the LS periodogram for a ROI around the Crab, periodic signals were detected only at the spacecraft orbital period of 1.6 hr and the standard LAT sky survey period of 3.2 hr, as expected. $Z_m^2$ test and H-test can clearly detect the periodic pulsation from the pulsar.

With all conventional methods failing to detect periodic signal from the very faint source, we utilized a temporal stacking method. By stacking different epochs into a phase-folded light curve, we may improve the signal-to-noise ratio (SNR) in the time bin of an emission pulse, and spatially isolate any signal from TVLM 513 using the {\it Fermi} likelihood analysis. Since the source is very faint, few if any signal photons are received in any individual time bin (some fraction of the rotation period), and the data are extremely noise dominated. For TVLM 513 we stack the data temporally by dividing it into time bins 1/10 the rotational period (roughly 12 min) using {\tt gtselect} with each of the time bin assigned with a phase. We then use {\tt gtselect} again to combine all event files with the same assigned phase in each time bin to create a phase-sorted dataset to run through the likelihood analysis pipeline described earlier. Thus we obtain the flux and TS in each phase bin, and a phase-folded light curve is created for the source. If the error of the flux in any phase bin is larger than the flux in that bin, of if the TS value is lower than 10, then the flux is replaced with a 95\% upper limit using the Upper\_Limits class provided by the $Fermi$ Science Tools. 


As mentioned, conveniently there is a 4FGL source within the ROI of TVLM 513 with a moderate GeV flux, 4FGL 1501. Since it flares randomly, 4FGL 1501 can be treated as a non-periodic source. While only separated by $0.2\degr$, the two sources should be readily separable in the likelihood analysis, especially at high energies. Since it is in the same ROI, we perform the identical likelihood analysis on 4FGL 1501 and use the results to verify those for TVLM 513.

\subsection{Results}
\label{subsec:tresults}

Following the methods mentioned in \S~\ref{subsec:temporalanalysis} and using the 7054.488 s period of \citet{harding2013}, TVLM 513 was not detected. 4FGL 1501 was detected in all phase bins (TS = 50-120), and no periodicity or pulse signature are evident. The lack of a detection from TVLM 513 at this point may be due to the relatively large uncertainty in the period used. Over many years, a pulsed signal could drift by several time bins. In fact, despite quoting a very high precision period,  \citet{wols2014} proposed a $\dot{P} = -1.13 \times 10^{-7}$ s s$^{-1}$, which would impose a significant drift not subsequently seen \citep{Lynch2015}.
For the 0.18 s uncertainty in the period used here, a putative pulsed signal would shift by one 11.76 min time bin after 3919 rotations, or just under a year. To check for any such drifting signal, we examined the phase-folded light curves in smaller time spans of roughly two years (9000 rotations). Despite the shorter exposures, 4FGL 1501 is detected again in most phase bins. Some marginal signal from TVLM 513 is seen in a couple of the time spans, reaching as high as TS = 22, and the detected fluxes, while weak, do appear to wander in phase indicating an imprecise period.

\begin{figure}
\epsscale{0.7}
\plotone{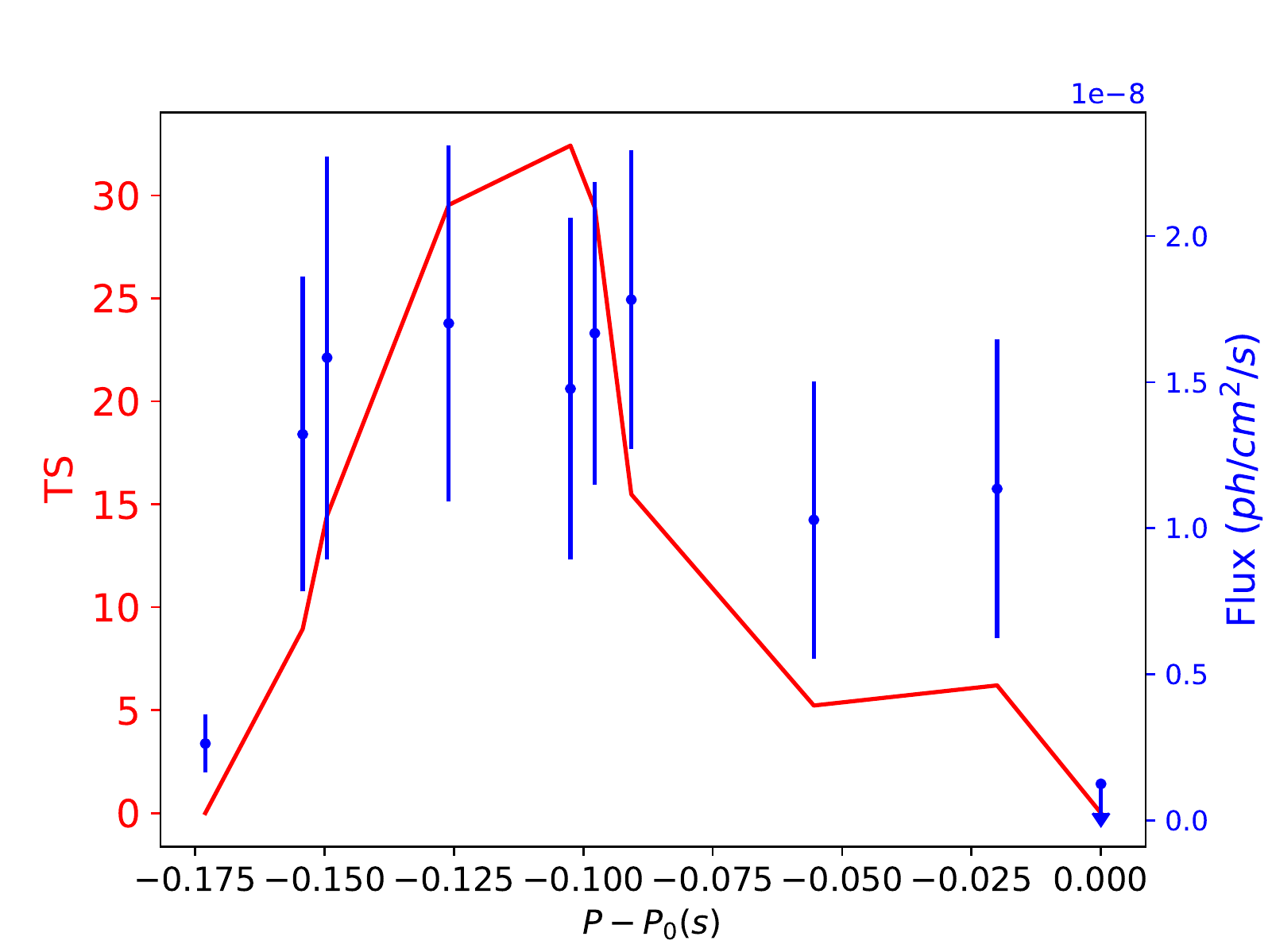}
   \caption{TS values (red) and flux (blue with errorbars) in phase bin 0.95 for different effective periods deviated from $P_0$.}
    \label{fig:shift}
\end{figure}

Rebinning the data from the start with a new period would have an unreasonable computational cost. Instead, we use the existing bins and shift the phase assignment one bin every several thousand rotations to effectively raise or lower the period within its uncertainty. For example, as discussed above, if the period were lower by the quoted uncertainty, the signal would shift to the next phase bin after every 3900 rotations. Shifting the phase assignment of a bin after 35,000 rotations brings the effective period down 0.020 s to 7054.468 s, the period determined by \citet{wols2014}. We tested a range of effective periods within the 0.18 s uncertainty of \citet{harding2013}, and found a consistent pulse at phase 0.95 that peaks in both flux and TS value for $P_{\gamma} = 7054.362$ s. Fig.~\ref{fig:shift} shows the changes in TS value and flux in phase bin 0.95 with respect to the period. Taking into consideration both the TS value and flux, there is a systematic rise in each to a peak for the optimal period at $P_{\gamma}.$

\begin{figure}
\plottwo{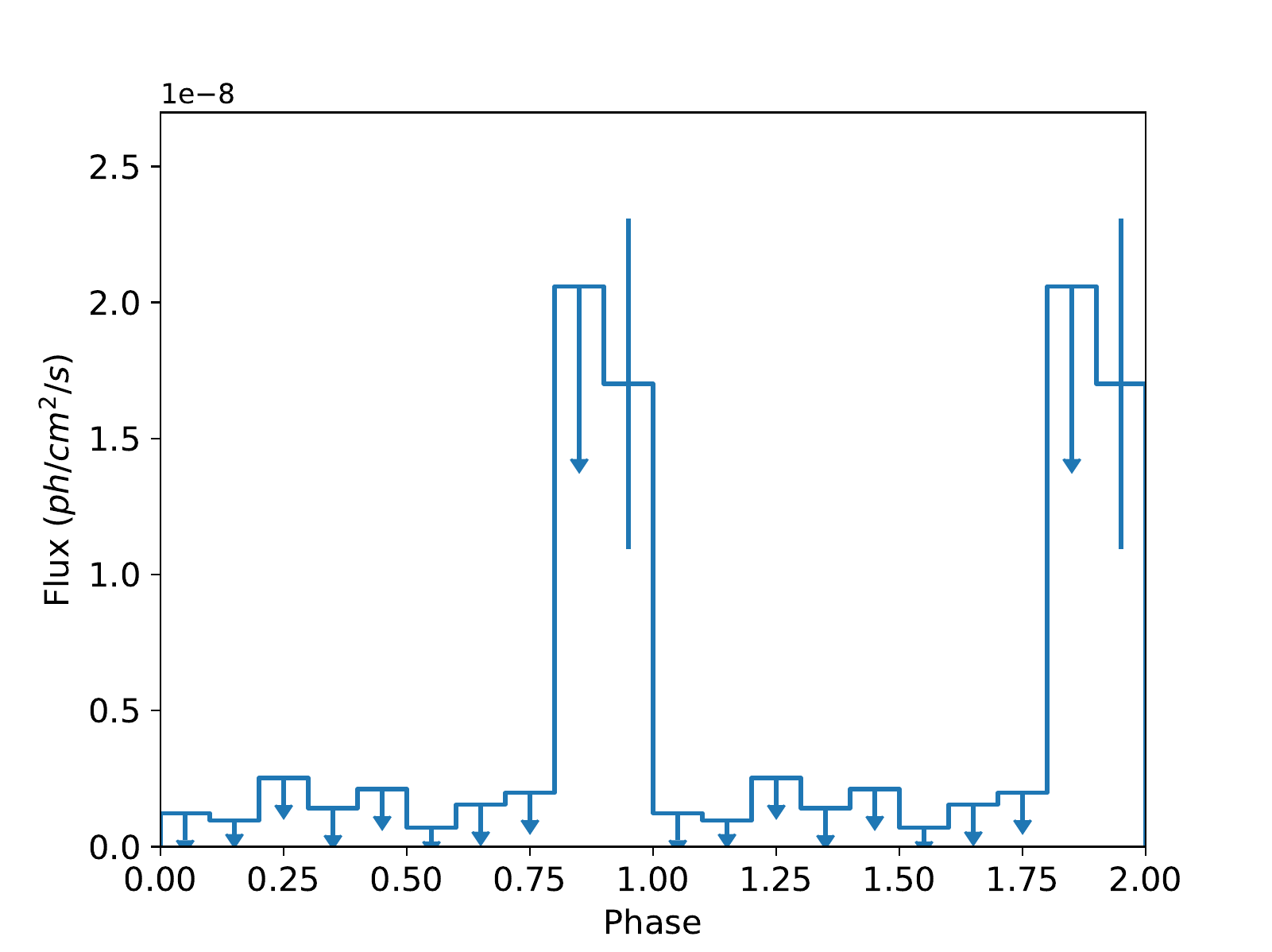}{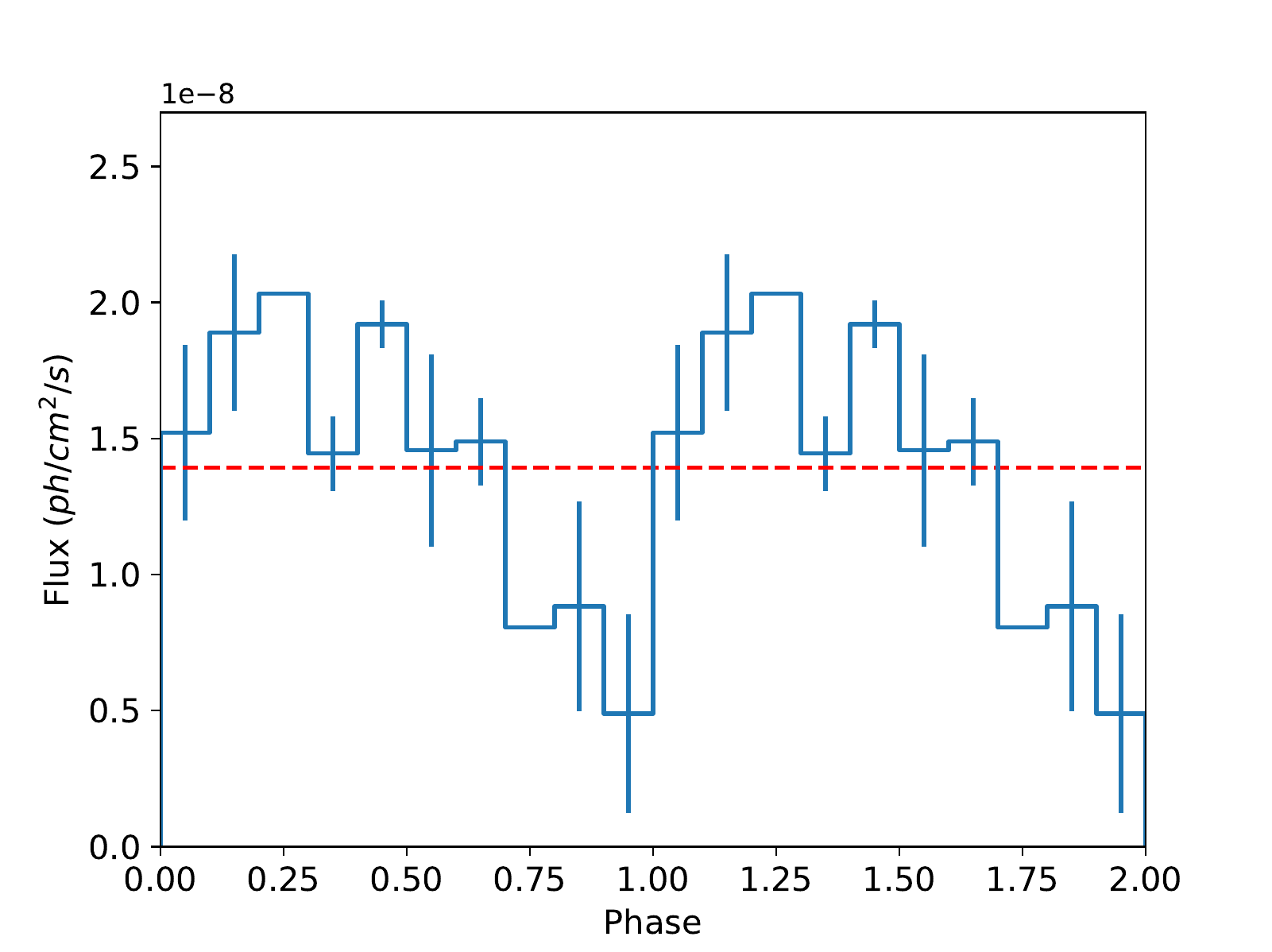}
\caption{Phase-folded light curves of TVLM 513 (left) and 4FGL 1501 (right) at the period of $P_{\gamma}$ = 7054.362 s. Upper limits were used in the light curve when TS in the bin is less than 25 or when the flux error exceeds the flux. The red line indicates the average flux of 4FGL 1501.}
\label{fig:12yrs}
\end{figure}

We rebinned the data using this new $\gamma$-ray period and ran through the entire likelihood pipeline from scratch. The resulting phase-folded light curves of TVLM 513 and 4FGL 1501 are shown in Fig.~\ref{fig:12yrs}. Again 4FGL 1501 is detectable in all phase bins and shows no periodic or pulsed signal. TVLM 513, on the other hand, has excess flux of $(1.7 \pm 0.6)\times10^{-8}$ ph cm$^{-2}$ s$^{-1}$ at a phase of 0.95 with a TS value of 29.5, and all the other phase bins have either low or undetectable fluxes. The power law index of the spectrum of the TVLM 513 pulse is $2.59 \pm 0.22$. A power law with exponential cutoff model also fits the data with a similar TS of 30.0, a photon index of $-2.54 \pm 0.24$, but the exponential cutoff energy is very poorly constrained. This TS value corresponds to a $5.0\sigma$ detection considering the 2 degrees of freedom of the power law model of the star \citep{wilks1938}. Taking into consideration the trials factor adjustment \citep{lyson2008} from scanning through 10 different periods, and using 10 phase bins per period, the significance becomes $4.0\sigma$ (see Appendix). It is noted that in phase bin 0.95 where the pulse of TVLM 513 is detected, 4FGL J1501 seems to have a minimum flux. 4FGL J1501 in this bin is well detected (TS = 30). Our statistical model described in the Appendix confirms that mis-assigning photons from a bright source to a nearby faint source does not result in any significant rise of TS value or flux of the faint source or any significant drop of TS value or flux of the bright source.

This $\gamma$-ray determined period agrees with the \citet{harding2013} value within the uncertainties, but differs significantly from the very precise value of \citet{wols2014}. However, they also mentioned that the pulse had ``jumps" into shorter periods, resulting in a large $\dot{P}$ as mentioned earlier.
Furthermore, \citet{Lynch2015} showed that the radio pulse of TVLM 513, while stable in phase, is complex in shape, showing features that span one to two-tenths of a rotation in phase over a few years. We therefore judge that our period is consistent with optical and radio measurements. Further, using a radio pulse arrival time from recent observations that overlap the {\it Fermi} data (M.\ Route, priv.\ comm.), we can readily compare the radio and $\gamma$-ray phase information. The phase of the radio pulse in the frame of the current analysis is 0.55. The radio pulse is known to precede the optical maximum by 0.41 in phase \citep{Lynch2015}, therefore our putative $\gamma$-ray pulse phase is $0.4\pm 0.05$ rotations from the radio, and closer in phase with the optical, differing by $0.1\pm 0.05$. These offsets in phase help shape the physical interpretation of the putative emission as described below.

\subsection{Control Fields}\label{subsubsec:tf}

We undertook a variety of tests to reduce the possibility of a spurious signal for TVLM 513 by repeating the analysis for apparently empty control fields, additional nearby 4FGL sources, and empty phase bins. To check for a systematic variation in TS and flux with effective period for TVLM 513, we analyzed an ``off" bin at phase 0.25 in the same manner as in Fig.~\ref{fig:shift}. The results indicate no equivalent systematic change in either quantity, and no instances of TS $>15$.
We modeled a mock source within the ROI of TVLM 513, but $0.2\degr$ on the opposite side of 4FGL 1501 to test whether the apparent signal may be some residual effect from modeling out the BL Lac, or some other spurious feature of the ROI. The phase-folded light curve of the mock source using $P_\gamma$ had no phase bins with TS above 6. We also analyzed 4FGL 1501 in the phase bin with the TVLM 513 pulse. Again, no significant systematic change in TS or flux develops for any particular effective period although the BL Lac is highly detectable in this phase bin for all periods.

\begin{figure}
\epsscale{0.9}
\plotone{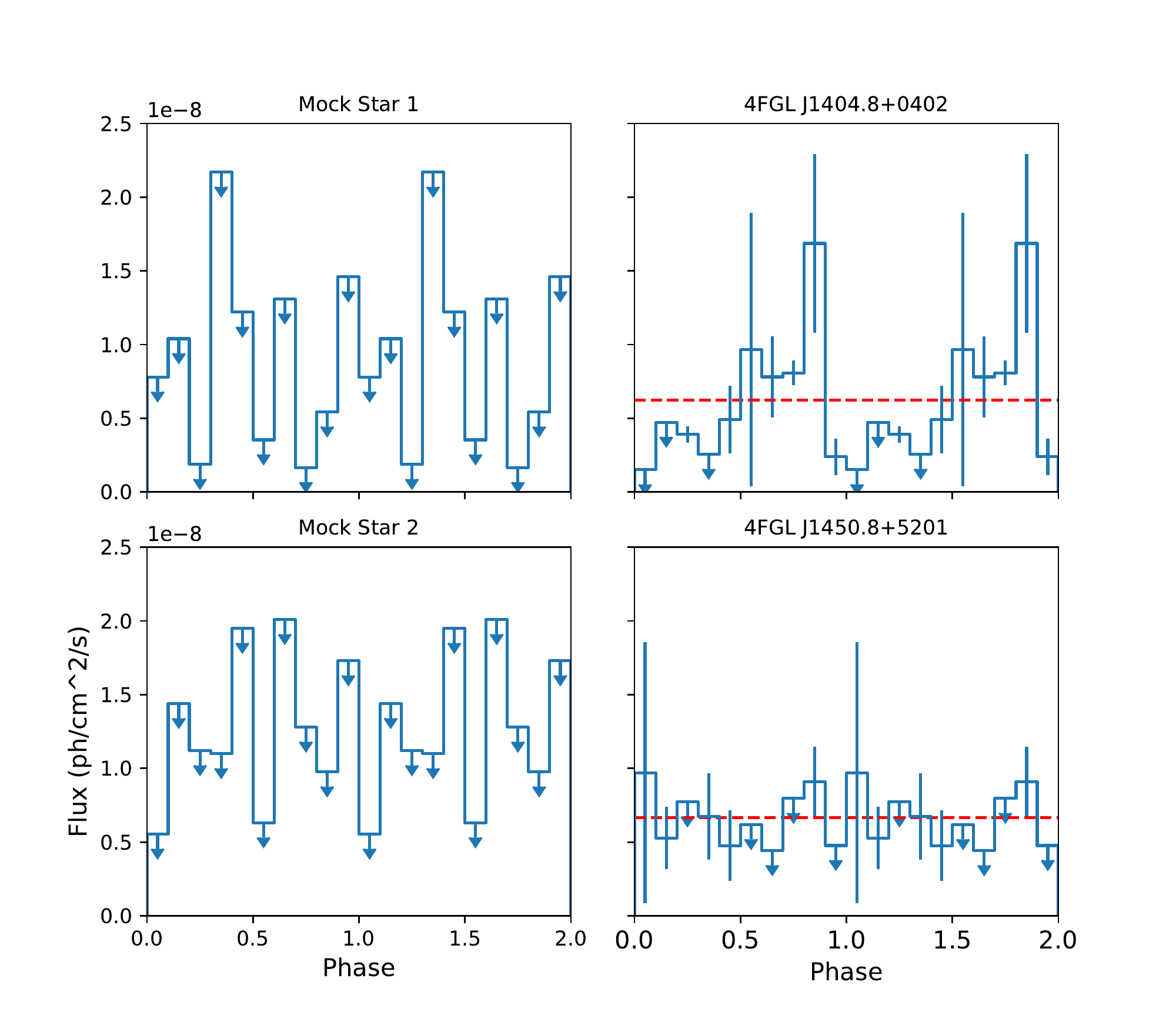}
\caption{Phase-folded light curves of two test fields. Light curve of the mock source in the first test field, marked as Mock Star 1, is shown in the top left panel and 4FGL J1404.8 in the top right panel. Light curve of the mock source in the second test field, marked as Mock Star 2, is shown in the bottom left panel and 4FGL J1450.8 in the bottom right panel. The red dashed lines represents the average flux of the sources over all 10 bins.}
\label{fig:falsepoz}
\end{figure}

We employed the exact same procedures described in \S~\ref{subsec:temporalanalysis} on two test fields, one east of and the other west of TVLM 513. These are at roughly the same Galactic latitude ($b\sim 60\degr$) so that all three fields have similar backgrounds. Both test fields contain faint 4FGL sources to serve as controls, and have the same ROI radius of $10\degr$. The first test field is centered on the quasar 4FGL J1404.8+0402, which is $23.2\degr$ away from TVLM 513. Mock star 1 is added $0.2\degr$ away from 4FGL J1404.8.
The second test field is centered on BL Lac 4FGL J1450.8+5201, $29\degr$ away from TVLM 513. Mock star 2 is added $0.2\degr$ away from 4FGL J1450.8.
None of these 4FGL sources have known periodicity at the period of interest. The phase-folded light curves of the 4FGL and mock sources, using $P_\gamma$, are shown in Fig.~\ref{fig:falsepoz}. Neither of the mock sources in the two test fields have obvious periodicity or pulsed signal, and the TS values of all phase bins are less than 13. The 4FGL sources also have no obvious periodicity or pulsed signal as expected.

\section{GeV Emission from Magnetically Active Stars} 
\label{subsec:auroral}

If the brightest source in the stack survey is at least a factor of 7 below the  $1.5\times 10^{-10}\ {\rm ph\ cm}^{-2}\ {\rm s}^{-1}$ sensitivity floor found in \S~\ref{subsec:smodel}, then its flux is $<2\times 10^{-11}\ {\rm ph\ cm}^{-2}\ {\rm s}^{-1}$, and the next brightest source is 30-50\%\ below that, depending on $\alpha$. The brightest $\gamma$-ray solar flares can reach a few $10^{-3}\ {\rm ph\ cm}^{-2}\ {\rm s}^{-1}$ \citep{Fermi_solarflares} and have a total energy as high as $10^{33}$ erg. Scaling this flux to the 8.4 pc average distance of the radio stars used here results in a flux of $10^{-15}\ {\rm ph\ cm}^{-2}\ {\rm s}^{-1}$. To reach the detection limit of our stacking survey would therefore require observing either $>10^4$ flares of this magnitude, or hundreds of superflares with energies above $10^{35}$ erg. The Evryscope flare survey \citep{howard2019} reached spectral types as cool as M4, for which they found flare rates of 12 and 0.14 yr$^{-1}$ for $10^{33}$ and $10^{35}$ erg flares, respectively. These rates would imply up to 9,000 solar-type flares from our survey, and about 100 superflares. While we have not observed any truly remarkable events, detecting stellar flares with significant ion acceleration activity appears to be just beyond the current stacking sensitivity limit of $Fermi$.



The GeV emission from these magnetically active stars is not yet very well understood, but non-thermal populations of accelerated particles are needed to emit photons with such high energies. Usually these same non-thermal particles are also associated with radio emission. Some of the UCDs have been studied extensively in radio wavelengths, and can exhibit periodic radio signals. The auroral emission from dwarf stars has been associated with the effect of either a close-in planet or acceleration at corotation breakdown \citep{Hallinan2015, Kao2016, Pineda2017}. The radio and $\gamma$-ray pulses from TVLM 513 are clearly also associated with its rotational period, and \citet{Kuznetsov2012} eliminated a close-in planet as a source of its radio emission. \citet{Lynch2015} found that an auroral model with pulsed radio emission tied to a magnetic loop footpoint fit the radio data for this star well, although simpler models of a dipole field and/or a long-lived active region are also quite valid \citep{Yu2011, Kuznetsov2012, Kao2016}. The distinction between flares or bursts and more persistent aurorae (or some combination) as the source of non-thermal particles is still unclear \citep{VilladsenHallinan2019}. We will assume the persistent active region configuration for the remaining discussion.

To constrain the proton population properties we model the $\gamma$-ray emission using the {\tt naima} package \citep{naima}. We input a $\gamma$-ray spectral energy distribution (SED) of the detected pulse using the results of the likelihood analysis from \S~\ref{subsec:tresults}, specifically the power law with exponential cutoff model. We assume a target proton density of $10^{14}$ cm$^{-3}$ which resembles the photospheric environment, though the normalization simply scales with the density. The resulting nonthermal proton spectrum has 
a spectral index of $\sim 2.6$ and an exponential cutoff energy around 30 GeV. The total proton kinetic energy is of order $10^{20}$ ergs. \citet{OhmHoischen2018} suggested that the total kinetic energy of these energetic particles makes up about 5\% of the total energy of the shock event, which implies a rather moderate total energy requirement. These energies are also consistent with, and yet do not exceed, the burst energies seen in radio flare stars \citep{VilladsenHallinan2019}. While the  uncertainties on these estimates are rather large due to the poorly constrained $\gamma$-ray spectral parameters, these results do not imply inordinate flare energies, luminosities or other unreasonable requirements to substantiate this GeV detection.

To accelerate the energetic protons, we may assume either powerful surface flares localized to a long-lived active region, and/or magnetospheric currents driven by the rotational shearing of a circumstellar plasma disk or other unknown mechanism. The nonthermal protons stream along converging field lines to an active region on the stellar surface. The radio-emitting electrons at TVLM 513 have been modeled similarly \citep{Lynch2015}, and we may compare those results with the implied proton densities from the $\gamma$-ray detection assuming a canonical proton-to-electron ratio of $\sim100$ \citep{schlickeiser2003}. For a 1 kG global field, the electron density to generate the quiescent gyrosynchrotron emission of these stars is about $10^5$ cm$^{-3}$ \citep{Lynch2015}. Integrating the proton spectrum to determine the proton density, the active region volume is the final unknown in the normalization, but may be constrained from the $\gamma$-ray spectral modeling. Given an active region surface area about 1\% that of the star, we estimate the implied active region thickness for a range of photon spectral index and cutoff energy values.
A minimum thickness, based on proton optical depths of 1--3 in the Sun \citep{Zhou2017} is likely to be 10s to 100s of km. Presuming an active region thickness no larger than its diameter sets an upper bound of about $10^4$ km. Even allowing for the large uncertainties assumed here for the photon and proton spectra, the detected pulse requires rather soft $\gamma$-ray spectra, 
and the proton spectrum mirrors these constraints. These parameters are very similar to the spectra of $\gamma$-ray solar flares \citep{Fermi_solarflares} and contrast with the hard proton spectra and TeV cutoff energies required by \cite{OhmHoischen2018} for flare star detections by current and future Cherenkov arrays.


%

\section{Conclusions} 
\label{sec:conclusions}

In this work, using 12 years of $Fermi$-LAT data, we report a  $\gamma$-ray pulse from the periodic M dwarf TVLM 513-46546. Its $\gamma$-ray period is $P_\gamma = 7054.362$ s, consistent with radio and optical observations. The pulse has a flux of $(1.7 \pm 0.6)\times10^{-8}$ ph cm$^{-2}$ with a power law index of $2.59 \pm 0.22$. The pulse has TS = 30, which taking into account the look-elsewhere effect and 2 d.o.f of the model, denotes a $4\sigma$ detection. A series of tests diminish the possibility of a false positive signal, including analyzing test fields, referencing to non-periodic {\it Fermi} catalog sources, scanning over different periods and statistical analysis of the detection.

The $\gamma$-ray pulse appears to be out of phase with the radio peak, but close in phase with the optical pulse, arguing for auroral emission tied to a magnetic flux tube footpoint on the stellar surface. While the details of the proton acceleration mechanism are not well understood yet, the energies and proton number densities required to generate the $\gamma$-ray pulse are far from excessive. In comparison to the quiescent radio emission, and assuming a reasonable proton-to-electron ratio and active region geometry, modeling the GeV pulse helps constrain the putative signal spectrum to be rather soft, similar to solar flare spectra seen with {\it Fermi}. TVLM 513 is potentially the first isolated, main sequence star other than the Sun detected in $\gamma$-rays. It should be noted that another candidate source, $\epsilon$ Eri, was reported with a tentative detection in $\gamma$-ray \citep{riley2019}, but the emission is likely from irradiated dust rather than magnetospheric acceleration of cosmic rays.

Also using nearly 12 years of $Fermi$-LAT data, we report sensitive upper limits of $1.5\times 10^{-10}\ {\rm ph\ cm}^{-2}\ {\rm s}^{-1}$ on the GeV flux of nearby flare stars by stacking 97 flaring M dwarfs detected in radio and/or x-rays. The upper limit from the stack null result is consistent with recent stellar flare surveys. The superflare rate would have to be factors of several higher than currently observed in order to exceed the sensitivity limit achieved in the stack.


\acknowledgments
The authors are extremely grateful to M. Kao, J. Villadsen, B. Chen, and J. Tan for very helpful discussions. The authors also thank the referee for their valuable input that helped greatly with this study. We are also grateful for the initial work on this project done by F. Rivera. This work was supported in part by the National Science Foundation under grants AST-1153335 and AST-1831412. Y. S. acknowledges support from the John P. McNulty Scholars Program for Leadership in Science and Math at Hunter College of the City University of New York.

\vspace{5mm}

\appendix

\section{Statistical Significance}\label{subsubsec:stats}

According to Wilks' theorem \citep{wilks1938} a TS value of 30 for 2 degrees of freedom (d.o.f) should correspond to a $5.0\sigma$ detection. However, since we searched through several periods and binned the data, this result could be reduced due to the ``look elsewhere effect" \citep{lyson2008}. The survival function of the $\chi^2$ distribution with 2 d.o.f at TS = 30 corresponds to a p-value of $p_0 = 3.0\times10^{-7}$. Given 10 periods and 10 phase bins, or 100 trials, the new p-value, calculated as $1-(1-p_0)^{100} = 3.1\times10^{-5}$, thus lowering the significance to $4.0\sigma$.

\begin{figure}
\epsscale{0.7}
\plotone{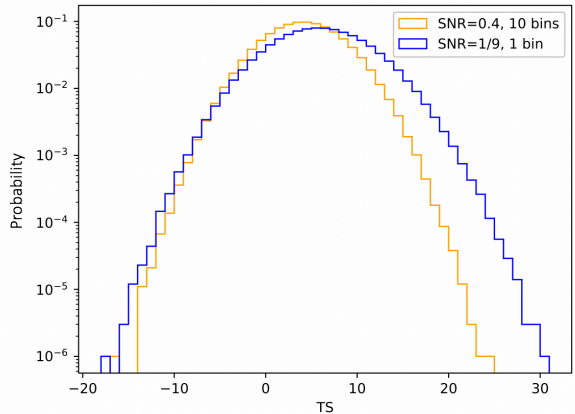}
\caption{Modeled TS distributions of TVLM 513 for $10^6$ trials as described in the text. The blue histogram indicates a minimal occurrence rate of a TS = 30 signal calculated from a SNR of 1/9. This SNR is therefore a lower limit requirement for a TS = 30 signal. The orange histogram represents the modeled TS values of the entire time series by combining 9 noise bins and one ``detection" bin with a SNR of 0.4. This SNR is therefore an upper limit for the pulse from TVLM 513 to remain undetected in the full (not phase-folded) likelihood analysis.}
\label{fig:ts_dist}
\end{figure}

To further quantify the potential false positive rate of a TS = 30 signal in one phase bin, we constructed a toy model to analyze the expected TS distributions and constrain the statistical significance. We assumed a Poisson noise level with an average of 2 counts per $0.1\degr$ pixel, consistent with the counts map of each of the phase bins of TVLM 513. We constructed two model ROIs, one containing noise, and one with a Gaussian star with a full-width at half-maximum of $0.5\degr$ combined with the noise. The stellar counts were also drawn from a Poisson distribution determined by the pulse SNR. Following the calculation in \citet{mattox1996}, we obtain the TS value of the stellar signal for $10^{6}$ trials. When the SNR $\ge 1/9$, TS values for the stellar signal larger than 30 emerge. The blue histogram in Fig.~\ref{fig:ts_dist} shows a distribution of TS values for SNR = 1/9, and TS exceeds 30 about once in a million trials. We therefore treat a SNR = 1/9 as a lower limit to the signal level of the stellar pulse. Lower SNR generally render the pulse undetectable, i.e., the TS distribution never exceeds 25 in $10^6$ trials.

While the phase-folded pulse signal from TVLM 513 may be detectable, the star was not detected in the spatial stack using the full unbinned counts. The pulse flux diluted by a factor of 10 falls well below the upper limit for TVLM 513 calculated in \S~\ref{subsec:temporalanalysis}. We apply our toy model again to determine the upper limit to the pulse signal, assuming it only occurs in one phase bin out of ten. The previously constructed star plus noise ROI is added to 9 more noise ROIs again drawn from a Poisson distribution with an average of 2 counts per pixel per bin. For a pulse SNR of 0.4, the resulting TS distribution does not exceed TS values of 25 in $10^{6}$ trials (orange histogram in Fig.~\ref{fig:ts_dist}). We therefore treat a SNR = 0.4 as an upper limit to the pulse signal of the star. The toy model TS value distributions assuming lower SNR or pure Poisson noise are very narrow and symmetric around zero, with TS $<15$ after many trials. This result is consistent with the likelihood analysis of the individual undetected ROIs in this work.

Given the low count rates, we also tested the sensitivity of the model TS distributions to the addition or removal of a small number of photons. This is particularly relevant given the proximity of 4FGL 1501 to TVLM 513. Artificially trading photons between overlapping sources resulted in negligible changes to the TS distributions. Specifically, assigning stray photons from a well detected source (SNR = 1.5) to an undetected source (SNR = 1/20) did not result in a spurious detection in 50,000 trials ($\Delta$TS $<1$).

\section{Temporal Model}
\label{subsec:tmodel}

We constructed a temporal model to simulate the light curve of TVLM 513 and gauge the significance of its possible detection. Only background noise and a stellar pulse are considered; potential contamination from other $\gamma$-ray sources is ignored. Within the same 12 year time range and $10\degr$ radius ROI we used for the data analysis in \S~\ref{subsec:temporalanalysis}, the output of {\tt gtmktime} was fed to {\tt gtbin} using the LC option.
Then {\tt gtexposure} was used to calculate the exposure for each time bin in the light curve. The sampling time for the temporal model was accumulated from this output for all time bins with non-zero exposure. We use the same bin size of 705.4362s as in \S~\ref{subsec:temporalanalysis}. The phase assignment is identical to the data analysis in \S~\ref{subsec:tresults} with respect to $P_\gamma$ in order to also replicate the {\tt gtselect} and period refinement processes. For the bins associated with the phase of 0.95, 
stellar signal combined with the background noise is assigned. All other bins are assigned only background noise. Signal and noise are both assigned in the same way as in the model described in the Appendix, so that TS values of the desired bins can be evaluated.

\begin{figure}
     
     \plotone{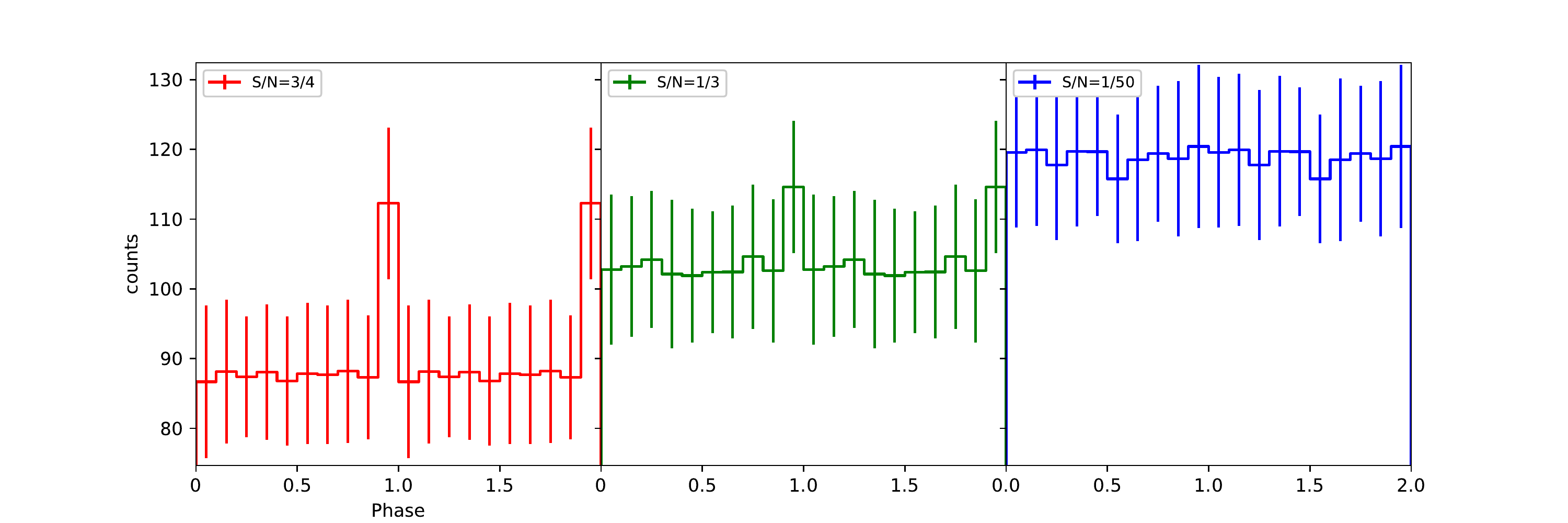}
     \caption{Modeled phase-folded light curves at different signal-to-noise ratios at period of $P_{\gamma}$. The counts in each bin are averaged over 100 trials, and the error bars are the standard deviation of the 100 counts. The left panel, with SNR = 3/4, represents a well detected source (TS $\sim 150$). The middle panel with SNR = 1/3, represents a marginally detected source like TVLM 513 (TS = 30). The right panel with SNR = 1/50 represents an almost pure noise light curve (TS $<4$). The total counts of the light curves are fixed, yielding the different normalizations of each. The error bars are the standard deviation of photon counts from 100 trials.}
     \label{fig:tmodel}
\end{figure}

The model background noise follows a Poisson distribution of average $1\times10^{-5}$, to recover the observed total photon counts per phase bin per pixel of $\sim 2$. This assures that the total number of counts here is at the same magnitude of observed counts in the ROI. We previously established that the reasonable range of SNR for the TVLM 513 pulse is between 1/9 and 2/5. 
We used a SNR of $1/50$ to represent pure noise, 1/3 for a marginally significant signal such as for TVLM 513, and a SNR of $3/4$ to represent a readily detectable signal. The phase-folded light curves from the model with these three SNRs are plotted in Fig.~\ref{fig:tmodel}. In the light curves, the counts of each phase bin are calculated by averaging results of 100 trials, and the error bars on the counts are the standard deviations of the 100 counts. 

\begin{figure}
\epsscale{0.7}
\plotone{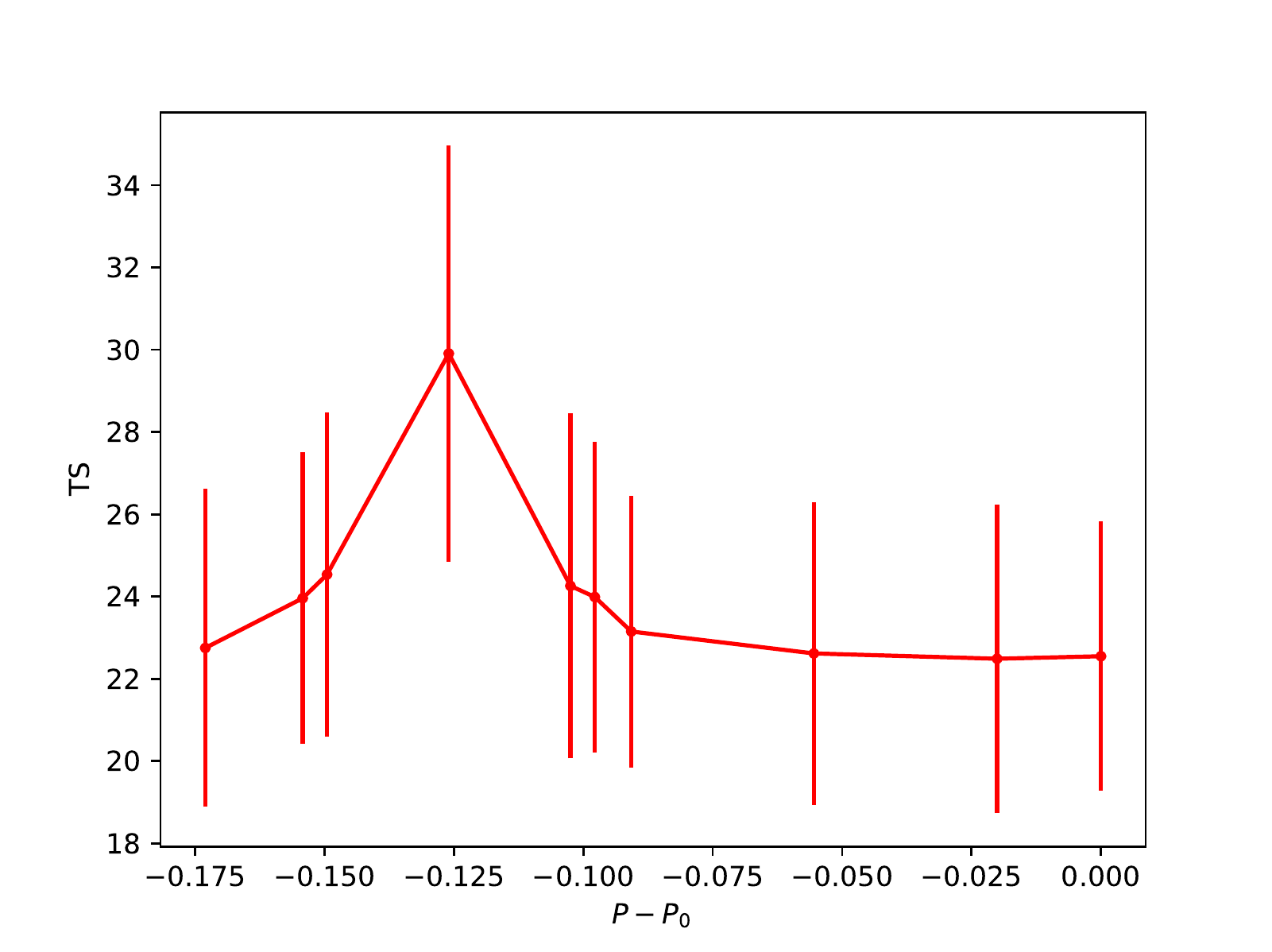}
     \caption{The modeled TS value in the phase bin 0.95 with respect to the period, using the temporal model with S/N = 1/3. Periods examined are the same as Fig.~\ref{fig:shift}. The error bars are the standard deviation of the 100 TS values calculated. At $P_\gamma$, the pulse has a TS near 30; at $P_0$, it drops below a $3\sigma$ signal.}
   \label{fig:tmodel_shift}
\end{figure}

The toy model light curves confirm that even a stellar signal as weak as $33\%$ of the noise is still discernible after stacking 12 years of data. The light curve with $S/N = 1/3$ is a close match to our results for TVLM 513, yielding TS near 30. 
The systematic increase in TS value towards the optimal period was also tested with the temporal model. We calculate the TS value of phase bin 0.95 at each period in Fig.~\ref{fig:shift}. Fig.~\ref{fig:tmodel_shift} shows that both the modeled counts and TS value peak at $P_{\gamma}$. TS values at phase bin 0.95 are calculated from the averaged counts over the 100 trials, and error bars are the standard deviation of the 100 TS values calculated from each individual trial. This result closely mimics that from the TVLM 513 period analysis. 

\end{CJK*}

\begin{thebibliography}{}

\bibitem[Abdo et al. (2009)]{abdo2009} Abdo, A. A., Ackermann, M., et al. 2009, \apj, 703, 1249


\bibitem[Acero et al. (2019)]{fermi2019} Acero, F., Ackermann, M., et al. 2019, arXiv, 1501.02003

\bibitem[Benz \& G\"udel (2010)]{BenzGuedel2010} Benz, A., O., \& G\"uedel, M., 2010, \araa, 48, 241B

\bibitem[Beran (1969)]{beran1969} Beran, R. J., Ann. Math. Statist, 40,1196


\bibitem[Buccheri et al.(1983)]{buccheri1983} Buccheri. R., et al., 1983, \aap, 128, 245

\bibitem[deJager et al. (1989)]{dejager1989} deJager, O. C., Swanepoel, J. W. H. \& Raubenheimer, B. C., 1989, \aap, 221, 180

\bibitem[G\"unther et al.(2019)]{Gunther2019} G\"unther M., N., et al., 2019, arXiv, 1901.00443

\bibitem[Hallinan et al.(2015)]{Hallinan2015} Hallinan, G., Littlefair, S.~P., Cotter, G., et al.\ 2015, \nat, 523, 568

\bibitem[Harding et al.(2013)]{harding2013} Harding, L. K. et al., 2013, \apj, 779, 101H

\bibitem[Henry \& McCarthy(1993)]{henry1993} Henry T.~J., \& McCarthy D.~W.~Jr.\ 1993, \apj, 106, 773

\bibitem[Howard et al.(2019)]{howard2019} Howard, W. S., et al., 2019, \apj, 881, 9H

\bibitem[Huber et al.(2012)]{Huber2012} Huber, B., Farnier, C., Manalaysay, A., et al.\ 2012, \aap, 547, A102

\bibitem[Kao et al.(2016)]{Kao2016} Kao, M.~M., Hallinan, G., Pineda, J.~S., et al.\ 2016, \apj, 818, 24


\bibitem[Kuznetsov et al.(2012)]{Kuznetsov2012} Kuznetsov, A.~A., Doyle, J.~G., Yu, S., et al.\ 2012, \apj, 746, 99

\bibitem[Li \& Ma (1983)]{lima1983} Li, T. P., \& Ma, Y. Q., 1983, \apj, 272, 317

\bibitem[Loh et al.(2017)]{loh2017} Loh, A., Corbel, S.; Dubus, G. 2017, \mnras, 467, 4462

\bibitem[Lomb (1976)]{lomb1976} Lomb, N. R., 1976, \apss, 39, 447L

\bibitem[Loyd et al.(2018)]{Loyd2018} Loyd, R.~O.~P., Shkolnik, E.~L., Schneider, A.~C., et al.\ 2018, \apj, 867, 70

\bibitem[Lynch et al.(2015)]{Lynch2015} Lynch, C., Mutel, R.~L., \& G{\"u}del, M.\ 2015, \apj, 802, 106

\bibitem[Lyson (2008)]{lyson2008} Lyons L., 2008,  Ann. Applied Stat., 2(3):887-915.

\bibitem[Mattox et al. (1996)]{mattox1996} Mattox, J. R., et al., 1996, \apj, 461, 396

\bibitem[McLean, Berger, \& Reiners(2012)]{mclean2012} McLean, M., Berger, E., Reiners, A. 2012, \apj, 746, 23

\bibitem[Mirzoyan (2014)]{mirzoyan2014} Mirzoyan, R., GRB Coordinates Network, Circular Service, No. 16238, \#1 (2014)

\bibitem[Ohm \& Hoischen(2018)]{OhmHoischen2018} Ohm, S. \& Hoischen, C. 2018, \mnras, 474, 1335

\bibitem[Omodei et al.(2018)]{Omodei2018} Omodei, N., et al, 2018, \apj, 865, L7

\bibitem[Pineda et al.(2017)]{Pineda2017} Pineda, J.~S., Hallinan, G., \& Kao, M.~M.\ 2017, \apj, 846, 75


\bibitem[Reid \& Hawley (2005)]{reid2005} Reid, I. N. \& Hawley, S. L. 2005, New Light on Dark Stars: Red Dwarfs, Low-Mass Stars, Brown Dwarfs (Berlin: Springer), p168

\bibitem[Riley et al.(2019)]{riley2019} Riley, A. H., et al, 2019, \apj, 878, 8R

\bibitem[Scargle (1982)]{scargle1982} Scargle, J. D., 1982, \apj, 263, 835S


\bibitem[Share et al.(2018)]{Fermi_solarflares} Share, G.~H., Murphy, R.~J., White, S.~M., et al.\ 2018, \apj, 869, 182

\bibitem[Shkolnik et al.(2009)]{shkolnik2009} Shkolnik, E., Liu, M. C., Reid, I. N. 2009, \apj, 699, 649

\bibitem[Schlickeiser(2003)]{schlickeiser2003} Schlickeiser, R. 2003., Cosmic Ray Astrophysics (Berlin: Springer), p473


\bibitem[Villadsen \& Hallinan(2019)]{VilladsenHallinan2019} Villadsen, J., \& Hallinan, G.\ 2019, \apj, 871, 214

\bibitem[Wilks (1938)]{wilks1938}Wilks, S. S., 1938, Ann. Math. Stat., 9, 60

\bibitem[Wolszczan \& Route (2014)]{wols2014} Wolszczan, A., \& Route, M., 2014, \apj, 778, 23

\bibitem[Yang \& Liu (2019)]{YangLiu2019} Yang, H., \& Liu, J., 2019, \apjs, 241, 29Y


\bibitem[Yu et al.(2011)]{Yu2011} Yu, S., Hallinan, G., Doyle, J.~G., et al.\ 2011, \aap, 525, A39

\bibitem[Zabalza(2015)]{naima} Zabalza, V.\ 2015, 34th International Cosmic Ray Conference (ICRC2015), 922

\bibitem[Zhou et al.(2017)]{Zhou2017} Zhou, B., Ng, K.~C.~Y., Beacom, J.~F., et al.\ 2017, \prd, 96, 023015

\end{thebibliography}
\end{document}